\documentclass[11pt]{article}

\usepackage{graphicx}

\newcommand{\ha}{\mbox{H$\alpha$}}
\newcommand{\hb}{\mbox{H$\beta$}}

\newcommand{\um}{\hbox{$\mu$m}}
\newcommand{\etal}{et al.\,}

\newcommand{\ergcmsqs}{\hbox{erg cm$^{-2}$ sec$^{-1}$}}

\newcommand{\arcmin}{\hbox{$^\prime$}}

\begin{document}

\def\eg{{\it e.g.}}
\def\etal{{\it et al.}}
\def\ni{\noindent}

\centerline{\bf \Large 
Deep slitless infrared spectroscopic surveys with HST/WFC3}

\begin{center}

\bigskip

a white paper for the working group on deep fields and large HST programs

\bigskip

August 31, 2012

\bigskip

Benjamin J. Weiner (Steward Observatory), bjw@as.arizona.edu

\end{center}
\medskip


\noindent 
{\large\bf Summary: }
HST is commonly thought of as an optical-IR imaging or UV-spectroscopy observatory.  However, the advent of WFC3-IR made it
possible to do slitless infrared spectroscopic surveys over an area significant 
for galaxy evolution studies ($\sim 0.15~\rm{deg}^2$).  Slitless
infrared spectroscopy is uniquely possible from space due to the reduced
background.  Redshift surveys with WFC3-IR offer probes of the 
astrophysics of the galaxy population at z=1-3 from line features, and the true redshift and spatial distribution of galaxies,
that cannot be done with photometric surveys alone.
While HST slitless spectroscopy is low spectral resolution, its high
multiplex advantage makes it competitive with future ground based
IR spectrographs, its flux calibration is stable, and its high spatial resolution allows measuring
the spatial extent of emission lines, which only HST can do currently
for large numbers of objects. A deeper slitless IR
spectroscopic survey over hundreds of arcmin$^2$ (eg one or more GOODS
fields) is one of the remaining niches for large galaxy evolution studies
with HST, and would produce a sample of thousands of spectroscopically confirmed galaxies
at $1<z<3$ to $H=25$ and beyond, of great interest to a large community of investigators.
Finally, although JWST multislit spectroscopy will outstrip HST in resolution and sensitivity, I believe it is critical to have a spectroscopic sample in hand 
before JWST flies.  This applies scientifically, to be prepared for the questions
we want to answer with JWST, and observationally, because JWST's lifetime is limited and a classic problem in targeted spectroscopy has been the turn-around time for designing surveys and for deciding which classes
of objects to target. This white paper is released publicly to stimulate 
open discussion of future large HST programs.

\section{Existing spectroscopic IR surveys and future niches}

Slitless spectroscopy with the WFC3-IR grisms on HST has a greatly increased survey speed
over what was attainable with NICMOS, and produces cosmetically clean data that can be processed through pipelines in large quantity.  This has enabled surveys covering most of several deep imaging fields (the CANDELS\footnote{http://candels.ucolick.org , PI S. Faber and H. Ferguson; Grogin \etal\ 2011} fields):
AGHAST\footnote{http://mingus.as.arizona.edu/$\sim$bjw/aghast/} in one field (PI B. Weiner) and 3D-HST\footnote{http://3dhst.research.yale.edu/ , Brammer \etal\ 2012} in four fields
(PI P. van Dokkum).  Together these cover
about 0.16 deg$^2$ with the G141 grism, wavelength range 1.1-1.65 \um, at a depth of 2 orbits/pointing, with an area of about 150 pointings
of 2 x 2 arcmin.  There is also a large parallel IR grism survey in effectively random fields, WISP\footnote{http://web.ipac.caltech.edu/staff/colbert/WISP/Home.html , Atek \etal\ 2010} (PI M. Malkan) and individual deeper grism pointings typically 6-8 orbit depth within the CANDELS fields as
part of supernova followup (PI A. Riess).

Extragalactic blank-field surveys have found a layer cake approach useful, with steps between wide, deep, and ultra-deep layers of order 1/4 the area and 4-10x the exposure time.
The 3D-HST + AGHAST surveys roughly correspond to the shallower, 2-orbit depth, ``wide'' layer of CANDELS, although they cover marginally less area and provide usable spectra only down to a limit 1-2 magnitudes brighter than the CANDELS-wide detection limit, since 2 orbits/pointing of spectroscopy does not go as deep as the CANDELS 2 orbits/pointing of imaging.

In the context of large community survey projects that might be carried out with either an allocation of director's discretionary time, future Treasury programs, or some combination,  
here I suggest the utility of a ``deep'' layer of grism IR spectroscopy.
This could cover of order 1-2 GOODS fields, or the CANDELS-deep parts of each GOODS, of order 100-200 arcmin$^2$: about 40 pointings
at depths of order 8 orbits/pointing in each of  G102 and G141 grisms,
providing spectral coverage from 0.8-1.65 \um, with HST's spatial resolution and good flux calibration.

Another obvious candidate for a large survey project is further 
deep field optical and IR imaging.
 In my opinion only, the niche for more blank-field imaging is not clear,
since at any layer (CANDELS-wide, CANDELS-deep, HUDF/HUDF09)
the expense of many hundreds of orbits would only increase area
by a factor of 2 or 3 at most.  Extremely deep exposure on small fields to detect high redshift galaxies is scientifically cutting-edge but 
also serves a small number of objects to a relatively small community of investigators.  Other less well-trod niches include imaging or spectroscopy behind lensing clusters for high-z galaxies; deep grism spectroscopy in one or a few fields for high-z line emission; and large
UV spectroscopy programs.  I expect these to be discussed in other
white papers.

\section{What can be done with IR grism surveys}

WFC3-IR G102 and G141 spectra cover 0.8-1.13 and 1.1-1.65 \um, and
in a deep-layer extragalactic survey, their chief yields are emission
line detections, out to z=1.5 for \ha, z=2.3 for [O III], and z=3.4 
for [O II].  
Absorption lines can be measured in grism spectra
but due to the low
resolution, detections of strong Balmer absorption in bright
post-starbursts are favored (eg van Dokkum \& Brammer 2010),
 and these are interesting but rare - we expect only of order 1/pointing.
The slitless spectroscopy yields many emission line spectra
per pointing, up to $>100$ redshifts in 2x2\arcmin, and can
produce redshifts for very faint galaxies.  

In 2-orbit data in low background fields, a moderately spatially extended
line is detected at 5$\sigma$ at flux of 5e-17 \ergcmsqs, and
8-12 orbit exposures increase S/N without reaching a systematic
floor or confusion limit, although multiple roll angles are needed
to resolve inter-object contamination.  The practical yield of similar data
is shown in Figures 1 and 2.  A new generation of ground
based IR spectrographs such as MOSFIRE are beginning to 
approach or better these limits in clean spectral regions, but still with 
lower multiplex and issues of sky subtraction and flux calibration.
HST IR spectroscopy 
covers some wavelength ranges inaccessible from the ground,
and yields accurate line and continuum flux calibrations; line
fluxes from ground based IR slit spectroscopy have been hard to
calibrate to better than a factor of 2 even in good conditions.

Science applications for IR spectroscopy, focusing on studying galaxies
and AGN at z=1-3, include:

$\bullet$ Star formation rates from \ha\ measurements, a
missing link between SFRs from far-IR and from the UV.  With
\ha\ and \hb, enabled by G102+G141 coverage at $0.8<z<1.5$,
extinction can be measured as a function of galaxy properties
such as SFR and stellar mass.

$\bullet$ Sizes of the star forming regions.  Slitless spectra
yield effectively narrow-band images of galaxies in their
emission lines.  These can measure the brightness of star
forming knots and whether z=1--2 galaxies form stars spread
out in massive disks, or centrally concentrated, as might be
expected in mergers, and can be compared to simulations of
the clumpiness of SF in high-z disks.  Such projects have begun,
especially for bright galaxies, with 3D-HST depth (Nelson \etal\ 2012),
but measuring sizes is difficult for fainter objects at low S/N and deeper data make a significant improvement.

$\bullet$ Line diagnostics of AGN and metallicity.  Grism spectra
of [O III]/\hb\ yield a probe of
AGN activity and of galaxy metallicity at $z \sim 2$
(Trump \etal\ 2011), only
studied in a handful of galaxies from the ground at these redshifts.  
AGN and low-metallicity galaxies
could be separated by reference to X-ray detections, galaxy mass, or
in some cases [S II]/\ha.  The grism offers a unique chance to measure
spatial variations in [O III]/\hb, and thus gradients on physical conditions or AGN emission, so far only possible in a stacked spectrum (Trump \etal\ 2011), but more feasible in deeper data.

$\bullet$ Strong emission lines from starbursting and low metallicity
galaxies.  Some high-z galaxies have emission lines so strong that
they produce unusual broadband colors, verified by grism spectra of [O
III] (van der Wel \etal\ 2011).  These likely represent short duration
events in low mass and low metallicity
galaxies and may be a clue to the nature of high-z galaxies, as well as an unusual and difficult-to-model population that could be a contaminant in color-selected high-z samples.

$\bullet$ Galaxy environments and close pairs from densely sampled
redshifts.  The dense sampling to faint magnitudes enables detection of close galaxy pairs for studies of the pair fraction, interactions, and galaxy satellite properties.  A dense spectroscopic redshift sample can measure galaxy environment in terms of local overdensity on Mpc scales more accurately than photo-zs, allowing studies of galaxy properties as a 
function of environment, and detection of groups.  Field size needs
to be close to a GOODS size to measure environments due to edge effects.

$\bullet$ Spectroscopic calibration of photometric redshifts for 
faint galaxies.  These extend to magnitude and redshifts poorly 
probed by existing spectra, critical for both galaxy evolution studies, and 
large dark energy programs relying on weak lensing.

$\bullet$ Ly-$\alpha$ emission lines at high redshift.  The
G102 spectra can detect Ly-$\alpha$ from $z=6-8$;
at this flux limit and area the expected sources are at a few times
$L^*$ (Ouchi \etal\ 2010) 
and the volume is $\sim 100/\phi^*$ per $\Delta z=0.5$, 
so one is sensitive
to rare bright sources.  G141 spectra could detect Ly-$\alpha$
at $z>8$, which would be of great scientific interest, but since
the Ly-$\alpha$  LF is unknown and likely to be fainter at high z, finding anything is highly uncertain.

The capabilities of grism surveys are shown in Figures 1 and 2, which
summarize a redshift catalog made from a grism observation
from CANDELS SN followup, overlapping the HUDF
\footnote{Data taken as part of PID 12099, PI A. Riess, Rodney \etal\ 2012; used courtesy of the CANDELS collaboration.}.  The catalog is
made by fitting redshifts to a G141 pointing of 6 and 4 orbit exposures at two roll angles, and using
a visual inspection GUI for quality control, with photometric redshifts
as an assist and to settle single-line identifications.  In this
2x2\arcmin\ field we found
144 good spectroscopic redshifts of which 87 are new, even in this
heavily observed field.  The catalog
shows that secure redshifts can be measured down to $H_{AB}=25.5$ and even a few at $H \sim 26$, although the yield is low. The real power of the grism comes in the range $0.9<z<2.4$ and $23<H<26$ where there
are many new redshifts; it reaches $\sim 2 $ mag below existing
spectroscopy.

\section{Survey design, data reduction, and cataloging practicalities}

Previous large community surveys, especially those done with 
director's discretionary time, have generally been imaging surveys
like the Hubble Deep Field or Ultra-Deep Field, and various
Treasury programs (GOODS, COSMOS, CANDELS, CLASH, PHAT, etc).
Exceptions include UV spectroscopy Key Project/Treasury programs and the 3D-HST Treasury,
which also  had well-defined proposal teams.  With imaging data,
typically many scientists inside and outside the proposal teams
have the experience to produce high quality reduced data and/or
to produce scientifically useful catalogs from reduced imaging data.

For large spectroscopic projects (IR and UV), there are several groups with the
capacity to reduce the data, but the reduction and cataloging ability
is not as widespread as for imaging.  For WFC3-IR grism data,
I have found that, after some effort, it is possible to produce 
good quality spectra in a useful format with a combination of
the aXe pipeline with close attention to its options and some custom code,
and to fit redshifts to the spectra. But the most time consuming 
step is quality control of the spectra and redshift estimates,
which so far has to be done by visual inspection with a GUI. 
I believe that other groups working on grism data have the same
issue, and in the ground-based DEEP2 survey (Newman \etal\ 2012) 
we also found that 
inspecting the $\sim 50,000$ spectra was one of the most labor-intensive
and arduous parts
of the survey. 

 In moderately deep grism data there are of order
500-600 spectra per pointing worth inspecting to $H_{AB}<26$, so 
a 40-pointing survey has 20-25,000 spectra to look at.  Clearly this is not
something one wants every user of the survey to have to do,
so if a survey were done with DD time it would be wise to provide 
resources and incentives to produce a public catalog and
data release, as would be expected of the proposal team for
an ordinary Treasury program.

\section{Why now with HST}

Any large infrared survey, imaging or spectroscopic, has to consider whether it will be instantly eclipsed by JWST observations, or whether it is valuable as a pathfinder for JWST.  JWST/NIRSPEC has greater sensitivity at higher resolution than HST/WFC3-IR grisms.
NIRSPEC requires assigning slits to objects, which causes it to fall below WFC3 in slit density by a factor of 4 or more: at $R\sim 1000$, roughly 70-100 slits in 3.5\arcmin\ x 3.5\arcmin, vs $\sim 140$ actual
measured redshifts in a single WFC3
2.1\arcmin\ x 2.1\arcmin\ pointing (Figure 1).  The efficiency of assigning slits is also usually closer to 50\% than 100\%.  Thus
to achieve equally dense sampling, redshift surveys with JWST/NIRSPEC will have to commit to visiting each tile of order 8 times or more, or revert
to low $R\sim 100$ spectra with NIRSPEC or NIRCAM/NIRISS grisms.
I believe this will make JWST redshift surveys expensive enough that
a moderate-area densely sampled survey will not be done immediately (while a very deep pointing for high-z galaxies will be).  

This leads to the crux of the problem: the limited JWST lifetime.  Experience with other surveys and telescopes shows that, for targeted
spectroscopy, there is a significant turnaround time as the users figure out how to optimize survey design and what targets are most interesting.  This is true even when the imaging for slit assignment already exists, as for example if JWST surveys a GOODS field.  Phasing this turnaround against a yearly proposal cycle may place JWST spectroscopy in a race
against mission lifetime.
Pre-existing low-resolution spectra will be invaluable to give us an idea of what high resolution spectra will reveal, and determine what objects are most critical for $R\sim 1000$ followup -- e.g. galaxies with unusual line fluxes or sizes, AGN candidates, X-ray or far-IR sources with interesting spectral properties; or compiling targets that sample the galaxy
population in a statistically controlled way.
A related example of this selection time delay in spectroscopy is the way that
the fraction of time proposed for and assigned to the IRS spectrograph on Spitzer increased significantly in later cycles.

\section{Conclusions}

Moderately deep WFC3-IR grism surveys can produce
large numbers of redshifts and emission line fluxes, of order 5000--10,000
redshifts in GOODS-sized fields, to faint IR magnitudes.  These
are extremely valuable for z=1-3 galaxy and AGN science, including
probing star formation rates, AGN detection, and spatially resolved
emission lines.  A deep layer of IR grism spectroscopy is a component
of deep-field extragalactic science that does not currently exist,
unlike deep imaging.  Determining whether such a large survey is better carried out through the regular proposal process or through DD time,
or both, and what are the required resources ito make it useful to the
community, are matters for the community, the working group on large
programs, and HST management to decide.

\bigskip

{\parindent 0pt
{\bf References:}

\bigskip

Atek, H. \etal\ 2010, ApJ, 723, 104

Brammer, G. \etal\ 2012, ApJS, 200, 13

Grogin, N.A. \etal\, 2011, ApJS, 197, 35

Nelson, E.J. \etal\ 2012, ApJL, 747, L28

Newman, J.A. \etal\ 2012, arXiv:1203.3192

Ouchi, M. \etal\ 2010, ApJ, 723, 869

Rodney, S.A. 2012, ApJ, 746, 5

Trump, J.R. \etal\ 2011, ApJ, 743, 144

van der Wel, A. \etal\ 2011, ApJ, 742, 111

van Dokkum, P. \& Brammer, G. 2010, ApJL, 718, L73

}

\bigskip

\begin{figure}[ht]
\centerline{
\includegraphics[width=5.5truein]{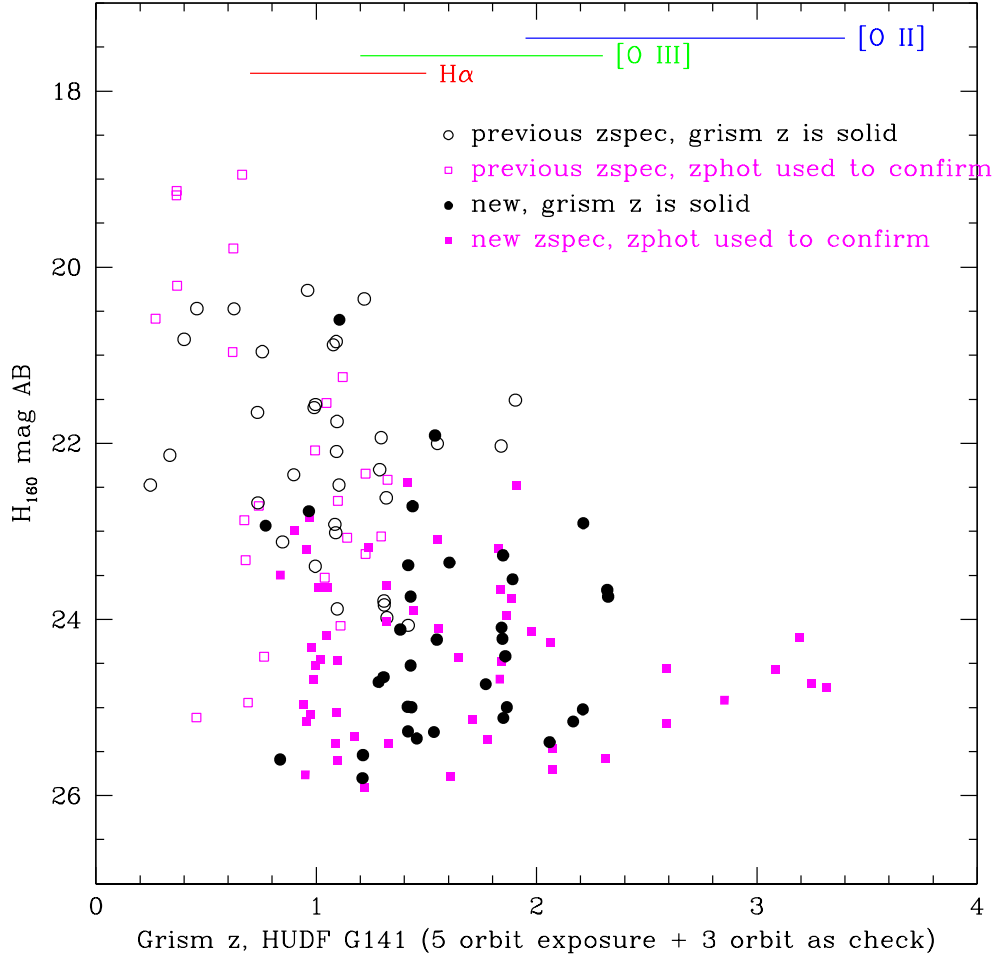}
}
\caption{Redshift vs. $H_{AB}$ magnitude from one WFC3 G141
  2x2\arcmin\ pointing
  in GOODS-S from CANDELS SN followup, split into 6+4 orbits at
  different roll angles.  We fit redshifts to spectra from the 6- and 4-orbit
  exposures separately, to $H<26$, and inspected visually for quality
  control. Horizontal lines at top indicate redshift ranges where
  strong nebular lines are detectable. Black circles are highest confidence grism redshifts (two
  or more features) and magenta squares are good quality (often a
  single line confirmed by a photo-z).  Open symbols had a
  spectroscopic redshift in the literature and filled symbols are new
  redshifts. There are 144 redshifts
of which 87 are new.  The grism yields many new redshifts in 
$0.9<z<2.4$ and $23<H<26$ .
}
\end{figure}

\begin{figure}[ht]
\centerline{
\includegraphics[width=5.5truein]{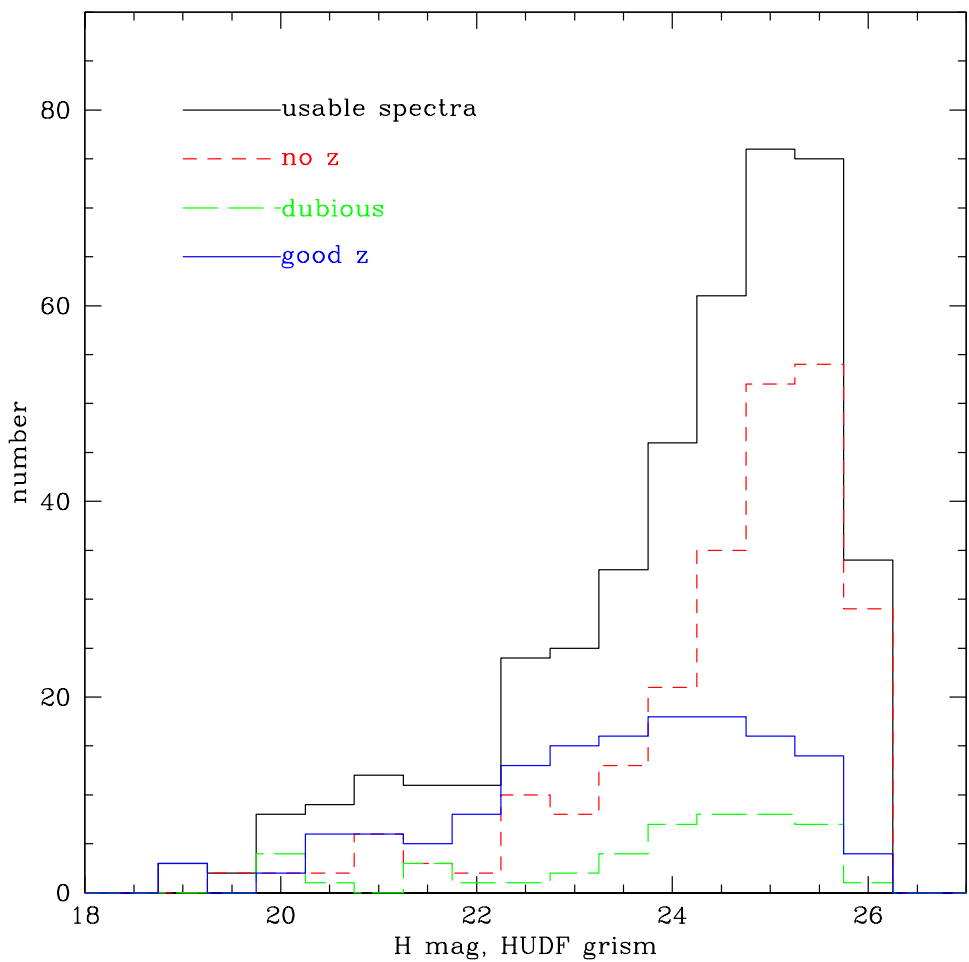}
}
\caption{Success rate for the WFC3 G141 pointing redshift catalog:
good redshifts, bad, and dubious/maybe are blue solid line, red dashed,
green long-dashed histograms.
The upper, black histogram of ``usable'' spectra gives the total number
excluding those spectra that fell off the detector or were heavily
contaminated by other sources.
Spectra were extracted down to $H=26$. The grism can yield
redshifts to very faint magnitudes although the success
fraction becomes low.
Bright galaxies for which the grism fails to get a redshift are
typically low-z without strong features at $\lambda>1.1$ \um.
}
\end{figure}

\end{document}